\documentclass[twocolumn,showpacs,preprintnumbers,amsmath,amssymb]{revtex4}

\usepackage{graphicx}
\usepackage{dcolumn}
\usepackage{bm}

\begin{document}
\draft
\newcommand{\lw}[1]{\smash{\lower2.ex\hbox{#1}}}

\title{Direct Extension of Density-Matrix Renormalization Group toward
2-Dimensional Quantum Lattice Systems: Studies for Parallel Algorithm,
Accuracy, and Performance}

\author{S. Yamada} 
\email{yamada.susumu@jaea.go.jp}
\author{M. Okumura}
\email{okumura.masahiko@jaea.go.jp}
\author{M. Machida}
\email{machida.masahiko@jaea.go.jp}

\affiliation{ CCSE, Japan Atomic Energy Agency, 6-9-3 Higashi-Ueno,
Taito-ku Tokyo 110-0015, Japan}
\affiliation{CREST(JST), 4-1-8 Honcho, Kawaguchi, Saitama 332-0012,
Japan}

\date{\today}

\begin{abstract} 

We parallelize density-matrix renormalization group to directly extend
it to 2-dimensional ($n$-leg) quantum lattice models. The
 parallelization is made mainly on the exact diagonalization for the 
superblock Hamiltonian since the part requires an enormous memory space
as the leg number $n$ increases. The superblock Hamiltonian is divided
into three parts, and the correspondent superblock vector is transformed
into a matrix, whose elements are uniformly distributed into processors. 
The parallel efficiency shows a high rate as the number of the states
kept $m$ increases, and the eigenvalue converges within only a few
 sweeps in contrast to the multichain algorithm.
\end{abstract}

\maketitle

The superfluidity achieved in atomic Fermi gas \cite{ColdAtomF} is quite
useful in studying strongly-coupled superfluidity. 
Very recently, such a success has intensively pushed experimentalists to
find another type of superfluidity, which emerges on strongly-correlated
2-dimensional (2-D) lattice system.
This is because the so-called ``optical lattice'' build in atomic gases
may offer a testbed to directly solve the Hubbard model and related
controversial issues in High-$T_{\rm c}$ cuprate superconductors in a
controllable manner \cite{toolbox}. 

So far, several computational approaches have been proposed in order to
 study strongly-correlated lattice fermions. 
Among them, three methods, i.e., the exact-diagonalization, the
density-matrix renormalization group (DMRG) \cite{White,DMRGreview}, and
the quantum Monte Carlo are widely employed as standard and established
ones. 
However, 2-D systems are too complicated for these methods to uncover,
and the ground-states in the most 2-D models are still open problems.  
In this paper, we therefore suggest a parallel algorithm to directly
extend DMRG to 2-D models. 

\begin{figure}
\includegraphics[scale=0.5]{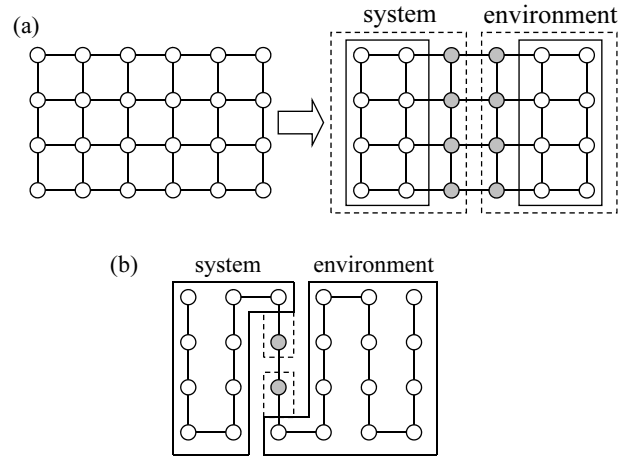}
\caption{\label{fig1} (a) A superblock configuration to directly
perform DMRG for 2-D lattice models. (b) The conventional configuration
using ``multichain'' algorithm to avoid enormous memory expansion. }
\end{figure}

The DMRG method, which was originally aimed for 1-D lattice model, can
be extended to 2-D models ($n$-leg models) as depicted in
Fig.~\ref{fig1} (a).
Then, the number of the states required in the direct algorithm is
roughly given as $16^sm^2$ ($4^sm^2$) for $s$-leg Hubbard model (the
$s$-leg Heisenberg model) per block, in which $m$ is the number of
states kept.  
Although the degree of freedom practically decreases with eliminating
irrelevant states, it is clear that a slight increment of the leg gives
rise to an exponential like growth of the state number. 
Thus, the direct extension has been limited within 2-leg \cite{NWS},
and the previous 2-D DMRG has adopted the so-called multichain algorithm
as depicted in Fig.~\ref{fig1} (b) since its memory space is basically
comparable to the 1-D case \cite{DMRGreview}.   
However, the multichain algorithm has difficulties in its convergence 
property and accuracy \cite{DMRGreview,Hager}.


The direct 2-D extension of DMRG guarantees high accuracy similar to 1-D
cases, although it requires an enormous memory space.  
Thus, if possible, it is valuable to parallelize the direct 2-D DMRG and
obtain a scalable code, which enables to raise the number of legs with
increasing computational resources. 
The present-day big supercomputers have a tera-byte order of memory.
We therefore claim that a scalable algorithm may be crucial in advancing
computational research on 2-D models.
We examine the parallel efficiency as well as the convergence property 
of 2-D direct DMRG on a parallel supercomputer Altix 3700Bx2 in JAEA.  
 

Let us explain the parallel algorithm.
In the direct 2-D DMRG as shown in Fig.~\ref{fig1} (a), a routine which
consumes most of the computer resources, i.e., memory and CPU time, is
the exact diagonalization of the superblock Hamiltonian $H$. 
Inside the routine, a major operation is the multiplication between the
Hamiltonian matrix and the vector, i.e., $Hv$. 
This is the most basic operation repeated over and again in the exact
diagonalization and DMRG.
In general, the parallelization of the multiplication between the sparse
Hamiltonian matrix \cite{note1} and the vector can be simply realized by
distributing the sparse matrix rowwisely. 
However, it is difficult to obtain a good load balance for the cases 
like the interacting 2-D lattice fermions, whose non-zero element
 distribution is not so regular.
In this paper, we, therefore, propose an alternative parallel strategy,
which transforms the superblock vector into a matrix form, and
distribute the matrix into processors. 

\begin{figure}
\includegraphics[scale=0.5]{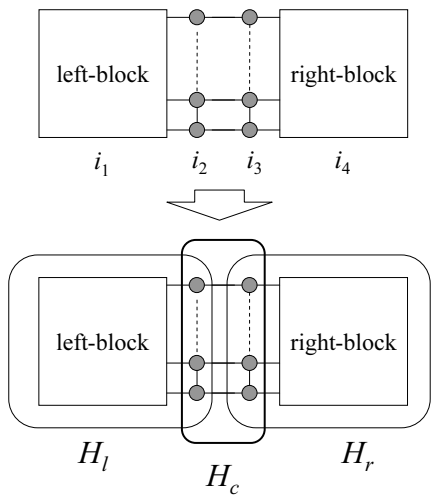}
\caption{\label{fig2} A superblock of 2-D direct DMRG, which is
composed of four blocks named ``block 1'', ``block 2'',``block 3'', and
``block 4'' from the left. See the text for the index $i_j$ (top panel).
$H_l$, $H_r$, and $H_c$ (bottom panel) denote the decomposed block 
Hamiltonians in the left, the right, and the central block,
respectively.}
\end{figure}

Let us write down the  algorithm.
Each block of the superblock is called ``block 1'', ``block 2'', ``block
3'', and ``block 4'' from the left, and the state of the ``block $j$'' is
represented as $i_j$  (see the top panel of Fig.~\ref{fig2}).
Then, the Hamiltonian matrix $H_{i_1 i_2 i_3 i_4 ;i_1' i_2' i_3'
 i_4'}(=H)$ is given by
\begin{eqnarray}
&& \hspace*{-5mm}H_{i_1 i_2 i_3 i_4 ;i_1' i_2' i_3' i_4'}\nonumber\\
&&=H_{i_1 i_2 ; i_1' i_2'}\delta_{i_3 i_4;i_3' i_4'}
+H_{i_3 i_4 ; i_3' i_4'}\delta_{i_1 i_2;i_1' i_2'}\nonumber\\
&&\hspace*{4mm}+H_{i_2 i_3 ; i_2' i_3'}\delta_{i_1 i_4;i_1'
i_4'},\label{dec} 
\end{eqnarray}
where $\delta_{i;j}$ is the Kronecker's delta, and $H_{i_1 i_2 ; i_1'
i_2'}$,\\ $H_{i_3 i_4 ; i_3' i_4'}$, and $H_{i_2 i_3 ; i_2' i_3'}$ are
the block Hamiltonian matrices in the left block, the right block, and
the central block, respectively. In the following, we express them as
$H_l$, $H_r$, and $H_c$ (see the bottom panel of Fig.~\ref{fig2}) for 
simplicity. 
Here, we put the $((i_3-1)m^2n+(i_4-1)mn+(i_2-1)m+i_1)$-th element of
the vector $v$, which corresponds to the state $|i_1i_2i_3i_4 \rangle$,
into an element $((i_2-1)m+i_1,(i_3-1)m+i_4)$ of a matrix $V$.
Then, the multiplication $H_l \delta_{i_3 i_4 ; i_3'i_4'}v$ and $H_r
\delta_{i_1 i_2 ; i_1'i_2'}v$ are rewritten as the following
matrix-matrix multiplications 
\begin{eqnarray*}
H_l \delta_{i_3 i_4 ; i_3'i_4'}v & \mapsto & H_{l}V,\\
H_r \delta_{i_1 i_2 ; i_1'i_2'}v & \mapsto & VH_{r}^T,
\end{eqnarray*}
where $H^T$ denotes the transpose of $H$.
Similarly, when the same element is put into the element
$((i_3-1)n+i_2,(i_4-1)m+i_1)$ of another matrix $V_c$, the multiplication
$H_cv$ is rewritten into
\begin{eqnarray*}
H_c\delta_{i_1 i_4 ; i_1' i_4'}v \mapsto H_cV_c.
\end{eqnarray*}
While the matrix $H_l$, $H_r$, and $H_c$ are sparse matrices, the
matrices $V$ and $V_c$ are complete dense ones since these are formed
by elements of the superblock vector $v$. 
This indicates that the parallel calculation for the matrix-vector
multiplication $Hv$ can be effectively executed by partitioning the
matrices $V$ and $V_c$.
This parallelization scheme has been successfully employed in the exact
diagonalization. 
The details of the parallel scheme and efficiency were reported in
 \cite{sc-2005,sc-2006}. 
By using this algorithm, one can extend DMRG to arbitrary $n$-leg model
as long as computation resources are unlimited.
In addition, the direct method has several advantages.
The application of the periodic boundary condition is not a problem at
all, and the extension to the time-dependent, the dynamical, and the
finite temperature DMRG \cite{DMRGreview} is straightforward. 
However, it should be noted that the increment of the ladder leg
enlarges not only the dimension of the Hamiltonian matrix but also that
of the density matrix.
Although the size of the density matrix becomes not so large, its
diagonalization needs $m$ eigenstates and its CPU time cost becomes
non-negligible with the leg increment. 
The parallelization in terms of the density matrix is another difficult
issue, since the size of the block diagonal matrices inside the density
matrix can not be predicted prior to the execution. The parallelization
should be adaptive to the dynamical change of the size.
Its algorithm and technique will be published elsewhere \cite{Yamada}.
In this paper, we restrict ourselves within the parallelization for the
Hamiltonian matrix operation, since it is the most primary issue for
2-D extension. The maximum leg sizes in this paper are 9 (its results
are not shown) and 5 for Heisenberg and Hubbard model, respectively,
due to the limitation of CPU resource \cite{note2}. 

\begin{figure}
\includegraphics[scale=0.5]{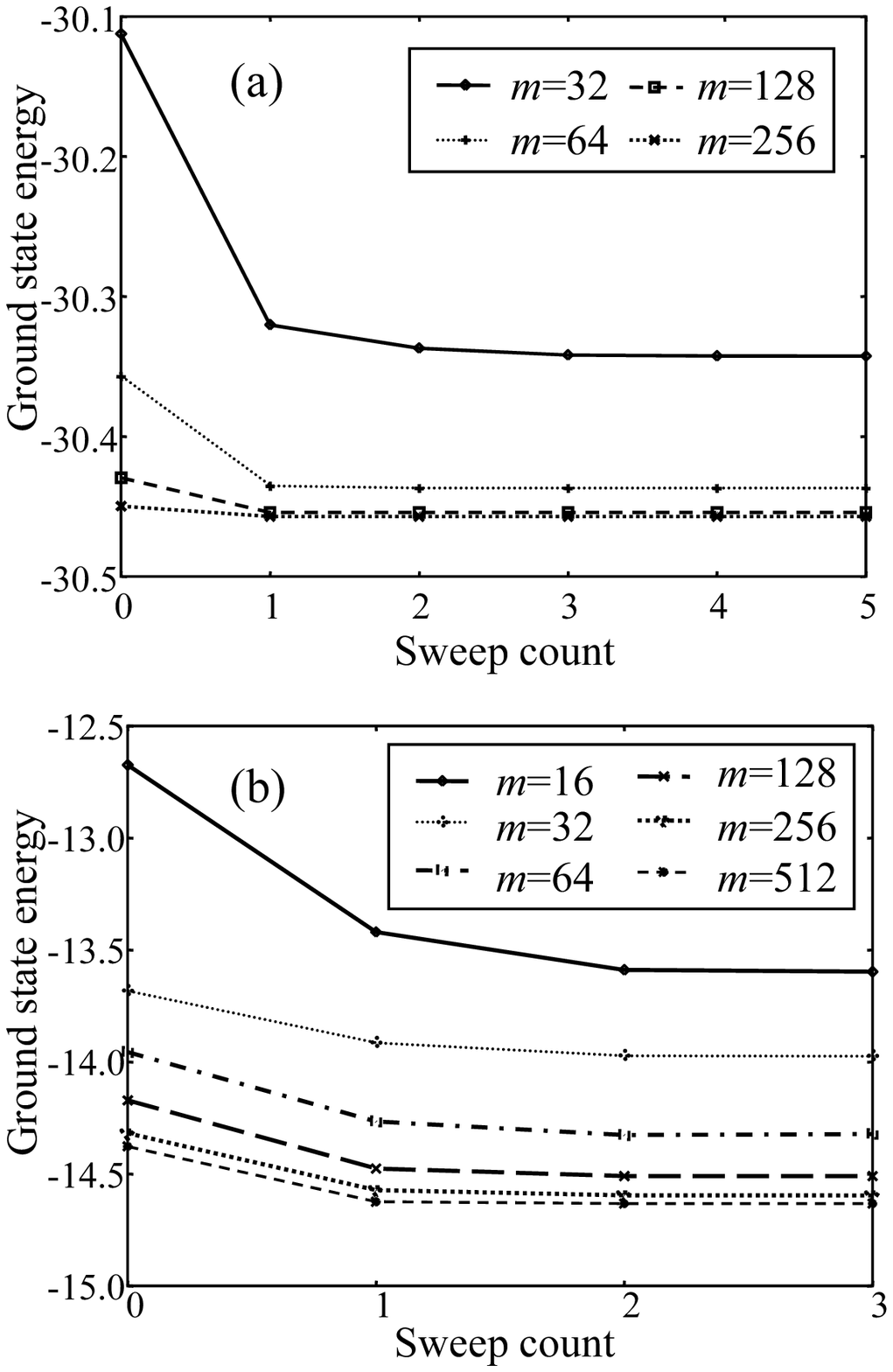}
\caption{\label{fig3}} The ground state energy vs. the sweep
counts for (a) $5 {(\rm leg)} \times 10$-site Heisenberg model and (b)
$3 \times 10$-site Hubbard model. $m$ is the number of states kept. 
\end{figure}

Let us present calculation results of the direct DMRG.
Figure \ref{fig3} (a) and (b) show how the ground state energy converges
with repeating the sweep for $5 ({\rm leg}) \times 10 $-site Heisenberg
model ($J_z=J_{xy}=1$) and the $3 \times 10$-site Hubbard model
($U/t=10$) with 28 fermions ($14\uparrow$, $14\downarrow$).  
The open boundary condition is applied to both models.
These results demonstrate that both models converge to their ground
state within once or twice sweeps \cite{3x6}. 
Moreover, the ground state energy sufficiently converges with $m \sim
128 $ and $\sim 256$ for Heisenberg and Hubbard model, respectively. 
These features clearly prove that the direct DMRG method is quite
excellent in the accuracy and the convergence properties in contrast to  
the multichain algorithm \cite{Hager}.

\begin{figure}
\includegraphics[scale=0.5]{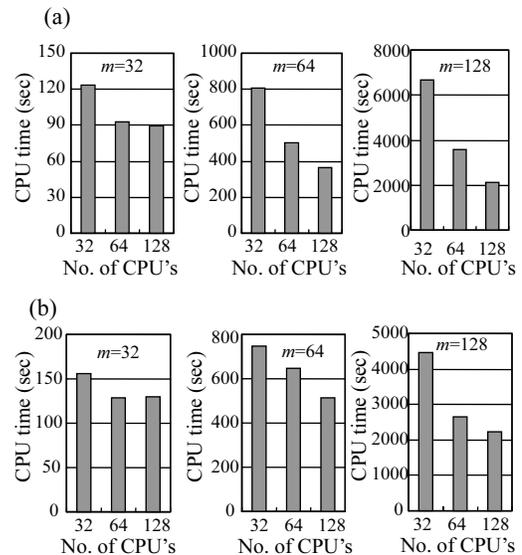}
\caption{\label{fig4} The CPU number dependence of the total CPU
time for (a) $7{\rm(leg)} \times 10$-site Heisenberg model on
Altix3700Bx2 with $m= 32$, $64$, and $128$. (b) The same dependence for
$4 \times 10$ Hubbard model.}
\end{figure}

Next, we present the performance of the direct DMRG method.
A test example is the two-dimensional $7\times 10$-site Heisenberg
model. 
Fig.~\ref{fig4} (a) shows how CPU time decreases with increasing CPU
number, and how the parallel scalability depends on $m$.
The latter effect is comprehensible through a comparison among $m=$32,
64, and 128. 
It is clear that the parallelization effect is improved when the number
of states kept $m$ increases. 
This is because the size of the Hamiltonian matrix to be diagonalized
grows with $m$, and the parallelized operation counts increases.
This behavior is common for $4 \times 10$-site Hubbard model with 38
fermions ($19 \uparrow$, $19 \downarrow$) as shown in Fig.~\ref{fig4}
(b). Thus, the direct 2D DMRG is a suitable application for parallel
computer, since the true ground-state exploration requires sufficiently
large $m$. 

\begin{figure}
\includegraphics[scale=0.9]{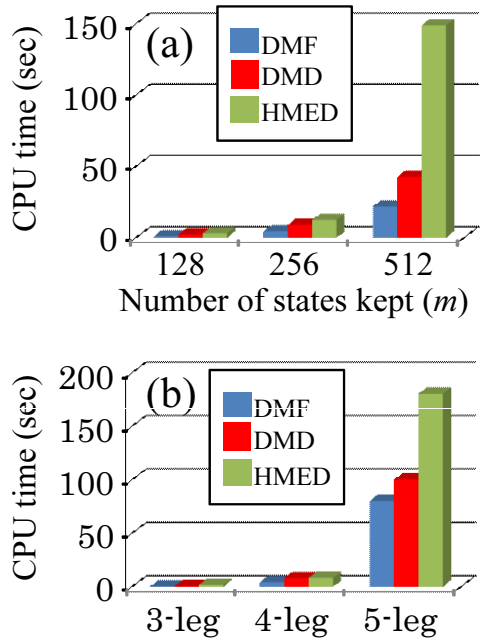}
\caption{\label{fig5} (a) The $m$ dependence of CPU cost
distribution for three main routines (see the text) of the present DMRG
code. The model is 3 (leg) $\times$ 10-site Hubbard model. CPU time of
each routine is averaged per step of DMRG. For every case, 128 CPU's on
Altix3700Bx2 are used. (b) The leg number dependence of CPU cost
distribution for the Hubbard model. In all cases, $m=64$, and 128 CPU's
are used.}
\end{figure}

Let us analyze details of the performance of the parallel direct
DMRG to discuss the feasibility of the present algorithm for larger
$m$ and leg systems. 
Figure \ref{fig5} (a) shows how CPU time of three main routines enlarges
with increasing $m$.  
The three routines are the density matrix formation (DMF), the density
matrix diagonalization (DMD), and the Hamiltonian matrix
exact-diagonalization (HMED).
The target model is 3-leg Hubbard model and 128 CPU's are used in all
cases for a comparison. 
One notices that the cost of HMED especially grows with increasing $m$. 
If HMED is not parallelized, then CPU cost for HMED is found to be too
huge. The parallelization for HMED is clearly crucial. 
Fig.~\ref{fig5} (b) is a leg number dependence of CPU time balance for
the three routines with $m=64$. 
Although HMED sustains the position as the heaviest routine on the
increment of the leg number, it is noted that the sum of DMF and DMD is
comparable to HMED for 5-leg model. 
These results indicate that CPU costs in terms of the density matrix
also becomes a bottleneck for larger leg cases. 

\begin{figure}
\includegraphics[scale=0.6]{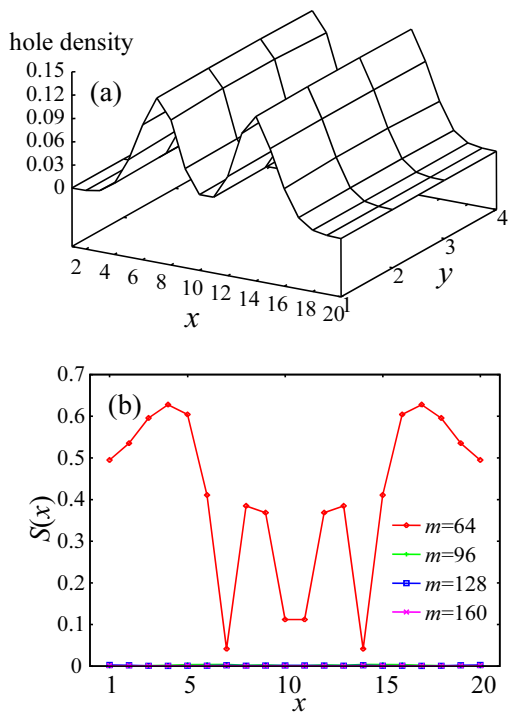}
\caption{\label{fig6} (a) The hole density profile for $4({\rm
leg})\times 20$-site Hubbard model ($U/t=10$) with 76 fermions ($38
\uparrow$, $38 \downarrow$ )and $m= 160$.  (b) The ladder direction
dependence of the maximum value of the local staggered spin density
along the leg direction for the same model with $m = 64, 96, 128$, and
$160$.}
\end{figure}

Finally, let us examine the validity of the ground state obtained by the
present direct DMRG. 
We pay attention to the Hubbard model with just below the half-filling.
The reason is that there is a controversial issue whether the stripe is 
the ground state or not \cite{Scalapino}.
We calculate spatial profiles of the hole density
\begin{equation}
h(x,y) = 1 - \langle \hat{n}_{x,y,\uparrow} + \hat{n}_{x,y,\downarrow}
\rangle \, ,  \label{hole}
\end{equation}
and the staggered spin density
\begin{equation}
s(x,y) = \langle \hat{n}_{x,y,\uparrow} - \hat{n}_{x,y,\downarrow}
\rangle \, ,
\end{equation}
with an open boundary condition for each direction.
Here, $\hat{n}_{x,y,\sigma} \, \, (\sigma=\uparrow, \downarrow)$ is 
the density operator and $\langle \cdots \rangle$ denotes the ground
state expectation value. 
One expects $s(x,y)=0$ for any local sites in the ground state of the
finite ladder Hubbard model, even if the hole density modulation
survives.  
This is a consequence of Lieb-Mattis theorem \cite{Hager,WhiteStripe}.
We calculate the ground state profiles of the $4 \times 20$-site Hubbard
model with 76 spins (38$\uparrow$, 38$\downarrow$) at $U/t=10$ with
varying from $m=64$ to $m=160$. 
Figure \ref{fig6} (a) shows the hole density profile for $m=160$, while
Fig. 6(b) presents $m$ dependence of the ladder direction profiles of
the maximum value of the staggered spin density along the leg-direction
given as  
\begin{equation}
 S(x)=\max_{y=1}^4|s(x,y)| \label{staggeed}.
\end{equation}
We point out that the present DMRG method rapidly converges
non-polarized pattern for the staggered spin density profile with 
increasing $m$, although the hole density one shows a stripe structure.
These results are different from those of the multichain algorithm
\cite{Hager}, in which an extrapolation is required to remove the
artificial profile of the spin density. 
To our knowledge, such a direct convergence is the first result 
in DMRG calculation of the 2-D Hubbard model. 

We developed a 2-D directly-extended code of DMRG.
We parallelized the exact diagonalization part by transforming the
superblock vector into the matrix and distributing the elements.
This parallel scheme becomes more effective as the number of states kept
$m$ increases.
In addition, we confirmed in the repulsive 2-D Hubbard model that the
stripe observable in the hole density profile is not artificial because
the spin-density modulation as its counterpart disappears with
increasing $m$ according to Lieb-Mattis theorem. 
We believe that the present direct 2-D DMRG will give a great impact on
the ground state exploration in atomic gas, solid state, and other
systems, by the future use of advanced parallel computers.

Two of authors (S.Y. and M.M.) acknowledge M.~Kohno, T. Hotta, and
H.~Onishi for illuminating discussion about the DMRG
techniques. M.M. also thanks Y.~Ohashi and H.~Matsumoto for the Hubbard 
model. 
The work was partially supported by Grant-in-Aid for Scientific Research
on Priority Area ''Physics of new quantum phases in superclean
materials'' (Grant No. 18043022) from the Ministry of Education,
Culture, Sports, Science and Technology of Japan. This work was also
supported by Grant-in-Aid for Scientific Research from MEXT, Japan
(Grant No.18500033).  



\begin{references}
\bibitem{note1} 
Generally, the block Hamiltonian matrix becomes sparse in 2-D models,
while it is not so in 1-D models.
\bibitem{note2}
We set 3 hours execution using 128CPU's on Altix3700 Bx2 as a 
resource limitation in this paper.
\bibitem{3x6}
For smaller ladder cases, e.g, 3$\times$ 6 Hubbard model, we confirm
that the converged eigenvalue shows an almost perfect agreement with
 that of the exact-diagonalization.
\end{references}
\end{document}